\documentclass[conference]{IEEEtran}
\usepackage{paralist}
\usepackage{pifont}
\usepackage{float}
\usepackage{bm}
\usepackage{amsmath}
\usepackage{comment}
\usepackage{cases}
\usepackage{bbm}

\usepackage{algpseudocode}
\usepackage{csvsimple}
\usepackage{array}
\usepackage{balance}
\usepackage{multirow}
\usepackage{mathtools}

\usepackage{mathptmx} 
\newcommand{\ignore}[1]{}
\usepackage[normalem]{ulem}
\usepackage[hyphens]{url}
\usepackage{fancyhdr}

\usepackage{enumitem}
\usepackage{caption}
\usepackage{graphicx}
\usepackage{verbatim}
\usepackage{color}
\usepackage{url}
\usepackage[bookmarks=false]{hyperref}
\usepackage[export]{adjustbox}
\usepackage[ruled]{algorithm2e}

\newcommand{\dname}{HLSDataset}

\begin{document}
%
\title{\dname: Open-Source Dataset for ML-Assisted FPGA Design using High Level Synthesis}

\author{\IEEEauthorblockN{ Zhigang Wei\IEEEauthorrefmark{1},
Aman Arora\IEEEauthorrefmark{2},
Ruihao Li\IEEEauthorrefmark{3},
Lizy John\IEEEauthorrefmark{4}}
\IEEEauthorblockA{\textit{The Laboratory for Computer Architecture, Department of Electrical and Computer Engineering} \\
\textit{The University of Texas at Austin}\\
Austin, United States \\
\IEEEauthorrefmark{1}zw5259@utexas.edu,
\IEEEauthorrefmark{2}aman.kbm@utexas.edu,
\IEEEauthorrefmark{3}liruihao@utexas.edu,
\IEEEauthorrefmark{4}ljohn@ece.utexas.edu
}
}
 \pagenumbering{gobble}

\maketitle


\thispagestyle{plain}
\pagestyle{plain}

\begin{abstract} \label{abstract}
Machine Learning (ML) has been widely adopted in design exploration using high level synthesis (HLS) for faster resource, timing and power estimation at very early stages for FPGA-based design. To perform prediction accurately, high-quality and large-volume datasets are required for training ML models. However, the current datasets used in this domain are proprietary or limited in use, and practitioners have to generate their own dataset to train HLS-related ML models. This paper presents a dataset for ML-assisted FPGA design using HLS, called HLSDataset. The dataset is generated from  widely used HLS C benchmarks including Polybench, Machsuite, CHStone and Rossetta. The Verilog samples are generated with a variety of directives including loop unroll, loop pipeline, and array partition to make sure optimized and realistic designs are covered. The total number of generated Verilog samples is nearly 9,000 per FPGA type. The dataset repository includes CSV (comma separated values) files containing both HLS and implementation metrics which can be easily consumed by ML model. We also include original C source code with directives, Verilog designs, post-HLS reports, post-implementation reports for each sample in the dataset, so that any metrics not present in the CSV can be easily extracted. In order to extend the dataset for future benchmarks, generation and extraction scripts are also provided. To demonstrate the effectiveness of our dataset, we undertake case studies to perform power estimation and resource usage estimation with ML models trained with our dataset. All the code and dataset are public at our github page\footnotemark. 
\footnotetext{https://github.com/UT-LCA/ML4Accel-Dataset/tree/main/fpga\_ml\_dataset}
We believe that HLSDataset can save valuable time for researchers by avoiding the tedious process of running tools, scripting and parsing files to generate the dataset, and 
enable them to spend more time where it counts, that is, in training ML models.
\end{abstract}
%
\IEEEpeerreviewmaketitle
\section{Introduction} \label{Introduction}

High-level synthesis (HLS) is able to convert software applications into FPGA hardware designs with different optimization strategies. It can greatly improve the productivity since hardware designers do not need to write low-level hardware description language (HDL) from scratch given an application written in a high-level language (HLL) like C, C++ or SystemC. 

While HLS greatly helps to reduce the effort for the software to FPGA implementation conversion, it is quite time-consuming, especially when large  design spaces need to be explored using various pragma settings.
This is a common usecase when designing application-specific optimized designs targeting FPGAs, for example, when designing FPGA based accelerators for ML applications.
Metrics such as resource usage and achieved clock frequency reported after HLS are estimates. To find the final metrics, the even slower implementation process (synthesis, place and route) is required. Even more efforts are needed to estimate power consumption accurately, since low-level simulation is required. For these reasons, efficient design space exploration targeting optimization of such metrics is hard. To address this challenge, machine learning (ML) based techniques are widely adopted to provide accurate resource usage and power estimation at early stage in HLS. S. Dai et al. \cite{hls-qor} uses Lasso linear model, XGB and artificial neural network (ANN) to calibrate the resource usage and timing results from HLS reports. Graph neural networks (GNNs) and HLS reports are used to predict performance in the work by N. Wu et al. \cite{hls-perf}. HL-Pow \cite{HL-pow} and PowerGear \cite{PowerGear} give solutions to predict power consumption using convolutional neural networks (CNNs) and GNNs respectively. E. Ustun et al. \cite{dsp-gnn} builds graph samples using the IR 
(intermediate representation) generated during HLS and use them as input to GNNs to predict operation delay.


\begin{figure}[hbt]
\centering
\includegraphics[width=0.8\linewidth]{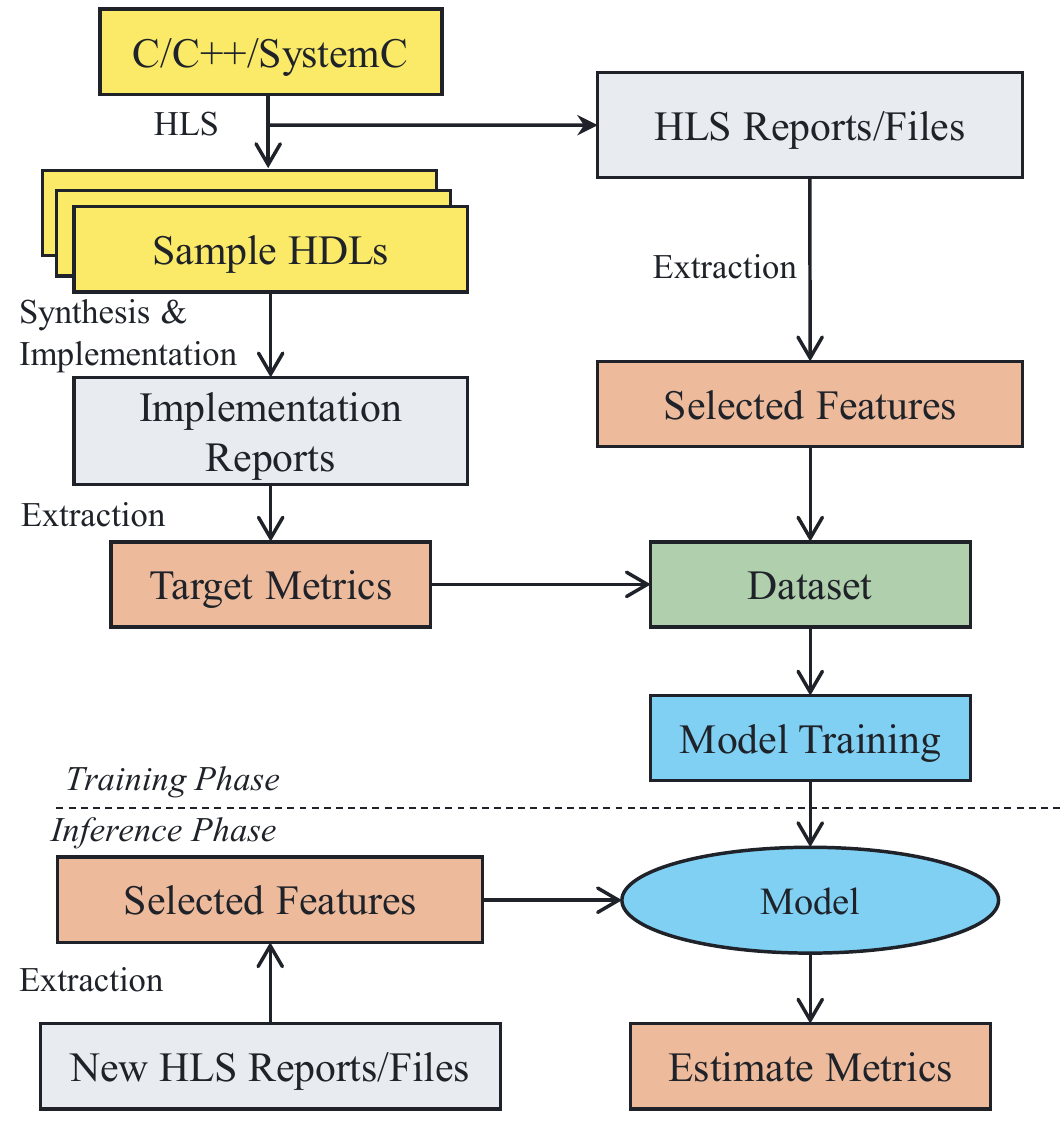}
\caption{The flow of general ML-based methods in HLS}
\label{fig:FLow_of_ML}
\end{figure}

The flow of general ML-based methods in HLS domain is shown in Fig \ref{fig:FLow_of_ML}. ML based methods can provide fast and accurate metrics estimation with HLS reports, however, extensive dataset is needed to train the models to produce acceptable results. To generate task-specific dataset in HLS domain requires lots of effort:
\begin{itemize}[leftmargin=*]
    \item Software source code should cover enough domains
    \item Source code should be well manipulated with HLS directives so that HLS optimization can be applied
    \item Varieties of optimization strategies need to be applied to the source code so that wide range of hardware designs can be generated
    \item Implementation is needed if the post-implementation metrics are the prediction goal
    \item Extensive scripting is required to extract the data from reports and preprocess before it can be consumed directly in ML models
    \item Significant computing resources may be needed for large number of tool runs to collect enough data
\end{itemize}

Researchers have to generate their own dataset, which can be extremely time-consuming because of the aforementioned reasons. Due to the different prediction goal and ML models, existing datasets are proprietary and not shareable or reusable.
However, there is an opportunity to reduce, and even eliminate, the redundant work for various researchers by creating a dataset that contains common usable information, allowing them to focus on training the ML models instead of generating the dataset.
We observe that resource usage reports, Intermediate Representation (IR) code, IR operator information, Finite State Machine Data path (FSMD) model from HLS are commonly used as the source of features. The resource utilization, timing information and power consumption values from post-implementation phase are the common metrics that researchers are interested to predict.




With the above observation, we propose HLSDataset: a well-curated open-source dataset for ML-assisted FPGA design using HLS. Our dataset can be used by a large subset of problems in this domain. 
The dataset currently contains nearly 9,000 Verilog designs per FPGA type, and two FPGA types are covered. To ensure diversity of designs, HLSDataset are generated from multiple applications across various benchmarks: Polybench \cite{Polybench}, Machsuite \cite{Machsuite}, CHStone \cite{CHStone} and Rosetta \cite{Rosetta}, and each application is tuned to generate a variety of hardware design samples.  Our dataset contains all necessary files and reports for every design (or, sample) so that features and target metrics can be easily extracted. In this paper, we describe the dataset, how it can be used, and showcase its utility by conducting two case studies. We expect this dataset to be widely usable and get even more useful with time through contributions by the FPGA research community.

Our contributions in this paper are as follows:
\begin{itemize}[leftmargin=*]
    \item Introduce HLSDataset and describe both the properties and usage of the dataset.
    \item Present the methodology how HLSDataset is generated. This methodology can be easily replicated to extend the dataset.
    \item Two case studies are conducted to demonstrate the effectiveness of HLSDataset.
\end{itemize}

Our dataset (including C code, Verilog code, CSV files, reports, and scripts) is open-sourced. 
The rest of this paper is organized as follows: Section \ref{Related work} summarizes the existing datasets and compares our dataset with them; Section \ref{Construction} illustrates the methods we use to generate HLSDataset; Section \ref{Properties} describes the contents of HLSDataset; Section \ref{Applications} gives a general overview of where HLSDataset can be used; Section \ref{Case studies} presents two case studies that use the dataset to successfully accomplish the prediction tasks, followed by a summary of this work and future work in Section \ref{Conclusion}.









\section{Related work} \label{Related work}
\begin{table*}[t]
\centering
\captionsetup{font=small}
\renewcommand{\arraystretch}{1.25}
 \begin{tabular}{|| >{\centering\arraybackslash}m{2.2cm} >{\centering\arraybackslash}m{1.3cm} >{\centering\arraybackslash}m{1.2cm}  >{\centering\arraybackslash}m{2.5cm} >
 {\centering\arraybackslash}m{2.5cm} >
 {\centering\arraybackslash}m{5.5cm} ||} 
 \hline
  \textbf{Work} & \textbf{\# Samples} & \textbf{\# Sources} & \textbf{Platform \& Abstraction level} & \textbf{Tools} & \textbf{Use Case in ML}\\ 
 \hline\hline
OpenABC-D~\cite{openabc_d} &  870,000 & 29 & ASIC RTL  & OpenROAD & Estimation of quality of a synthesis recipe \\
\hline
CircuitNet~\cite{circuitnet} & 12,960 & 6 & ASIC Physical Design  & Synopsys DC & Congestion prediction, DRC violation Prediction, IR drop prediction \\
\hline

Dai~\cite{hls-qor} & 1,300 & 65 & FPGA HLS & Xilinx Vivado & Quality of Results Estimation on one FPGA   \\
\hline
MLSBench~\cite{mls_bench} & 6,000 & 30 & FPGA HLS  & Xilinx Vivado & NA  \\
\hline
Spector~\cite{spector} & 8,300 & 9 & FPGA HLS  & Altera OpenCL SDK & NA  \\
\hline

Ours  & 18,876 & 34 & FPGA HLS  & Xilinx Vivado & Power Estimates, resource and timing estimation, operation delay estimate, cross-FPGAs studies, and more \\ 
\hline
\end{tabular}
\caption{Comparing HLSDataset with prior open-source datasets for training ML models for chip design}
\label{table:comparision}
\end{table*}

The success of ML-based models depends on well-curated datasets.
There are a few datasets for training ML models to assist in chip design problems in the ASIC domain.
OpenABC-D \cite{openabc_d} from NYU is a large-scale, labeled dataset produced by synthesizing open source designs with an open-source ASIC logic synthesis tool. This dataset can be used in developing, evaluating and benchmarking ML-guided logic synthesis
but is applicable to a very small subset of problems i.e. prediction of ASIC synthesis results.  
CircuitNet \cite{circuitnet} is another open-source dataset targeted for three prediction tasks in backend ASIC flows - congestion prediction, DRC (Design Rule Check) violation prediction, and IR drop prediction. It contains more than 10000 samples (in form of 2D image-like data) obtained by running open-source RISC-V designs through commercial EDA tools. This dataset is applicable to only a few physical design problems.

For FPGA HLS design flow, which is the focus of this paper, there are a few open-source datasets as well.
Dai et al. \cite{hls-qor} have open-sourced a dataset that is applicable to prediction of resource usage and delay (or frequency) for FPGAs  from high-level applications written in C. 
The dataset is generated by using applications from suites such as CHStone, Machsuite and Rosetta, and the Vivado tool chain from Xilinx/AMD.
This dataset is restricted to use only in estimation of resource usage and timing for FPGA, and contains only limited data.
The data provided is only for 1 FPGA device, implying that this dataset can not be used for cross-FPGA predictions.

MLSBench \cite{mls_bench} is an open-source dataset generated from 17 C/C++ and 13 SystemC benchmarks using Xilinx Vivado HLS tool flow. 
The C sources to generate the designs are from S2CBench \cite{S2CBench}, CHStone \cite{CHStone} and MachSuite \cite{Machsuite}. 
The dataset contains only log files and reports generated from Xilinx Vivado HLS tool flow, but without directly consumable features, labels and RTL codes. Also, this dataset is limited to only one FPGA. Therefore, MLSBench is hard to extend and quite limited in ML usage.

Spector \cite{spector} is a benchmark suite that contains applications written in OpenCL. 
The authors run the benchmarks through Intel OpenCL SDK to generate 8300 hardware designs targeted for Intel FPGAs.
In addition to just the benchmarks, several metrics for each design sample (based on compilation using Intel OpenCL SDK) are also provided.
The focus is on HLS tool flows and design space exploration.

Table \ref{table:comparision} compares the various properties of these datasets. We show the number of samples contained in the dataset and number of sources used for generating the dataset.
These datasets generally cater to limited usecases (eg: physical design prediction in ~\cite{circuitnet}, or RTL synthesis quality prediction in OpenABC-D \cite{openabc_d} or resource usage prediction in Dai et al.\cite{hls-qor}). Some need further expansion and curation to be readily usable by others.
Retargeting the few available datasets for a new ML model requires significant manipulation and augmentation.
So, researchers often generate their own dataset every time they want to solve a new problem.
In this process, they have to rerun tool flows to generate reports and then write scripts to parse those reports repeatedly. 
This motivates us to develop a dataset that is retargetable, versatile and robust, so that researchers do not need to replicate the tedious process of generating the dataset.

We focus on developing a dataset for predictions from applications written in high-level languages (HLLs) because high-language models of  applications are available in early stages of development of customized designs such as application-specific accelerators. 
In other words, we focus on prediction at the HLS level.
Predicting at the HLS level provides the most benefit in design space exploration.
We present HLSDataset, an open-source dataset for ML-Assisted FPGA Design for HLS. 

\section{HLSDataset Construction} \label{Construction}

Table \ref{table:suumary of HLSDataset} gives a general overview of our HLSDataset. 
We use HLL sources belonging to various application domains such as multimedia, arithmetic, signal processing and machine learning, from multiple popular benchmark suites such as Polybench \cite{Polybench}, Machsuite \cite{Machsuite}, CHStone \cite{CHStone} and Rosetta \cite{Rosetta}.
 Xilinx Vivado/Vitis tool chains are used for HLS and implementation.
Two FPGAs are used: ZU9EG and XC7V585T. We plan to expand the dataset to include more FPGAs, including Intel FPGAs.
One target frequency of 100 MHz is used. We are working on using more target frequencies as well.

\begin{table}[hbt]
\centering
\captionsetup{font=small}
\renewcommand{\arraystretch}{1.25}
\begin{tabular}{p{2.3cm}  p{5.7cm} } 
\hline
\textbf{Category} &\textbf{Details}\\
\hline
Num samples         &  18,876 \\
Num applications    &  34 \\
Application sources       &  Polybench, Machsuite, CHStone, Rosetta \\
FPGAs       &  ZU9EG, XC7V585T \\
Clock frequency & 100MHz \\
Domains          & Multimedia, Arithmetic, Signal processing, ML \\
Size           &  50 GB \\
Machines         &   9 16-core Intel Xeon 5218 2.3GHz 384 GB RAM\\
Time    &  More than 1,500 hours \\
Tools  &  Xilinx Vivado/Vitis \\
\hline 
\end{tabular}
\caption{General overview of HLSDataset}
\label{table:suumary of HLSDataset}
\end{table}

\subsection{C source code manipulation}
Verilog designs generated from C benchmarks are highly dependent on HLS directives, pragmas and the target clock frequency. For generating our dataset, we focus on the design space of \textit{loop unroll}, \textit{loop pipeline} and \textit{array partition}. Loops in C code need to be labelled so that loop unroll and loop pipeline can be applied to generate efficient designs. 
Machsuite and Rosetta are already well-written with HLS directives, and we directly use their code for our dataset generation. 
We manipulate the Polybench and CHStone source code with HLS directives. 

\subsection{Auto-generation of Tcl scripts}

The scope of generated Verilog designs can be huge, since the factors for \textit{array partition} and \textit{loop unroll} can vary greatly. The number of generated designs is determined by the number or the dimension of the factors we want to explore in our dataset. However, manually writing every Tcl script (Xilinx Vivado/Vitis tools use a Tcl script based interface), which is used to tune HLS solution for the generation of Verilog designs in our dataset, is time-consuming and unrealistic. In order to generate designs more efficiently, we write a template Tcl script for every C source code and a script to parse it. The script will auto-generate Tcl files which can be directly used by the HLS tool.

\begin{figure}[hbt!]
\centering
\setlength{\fboxsep}{2pt}
\fbox{\includegraphics[width=1\linewidth]{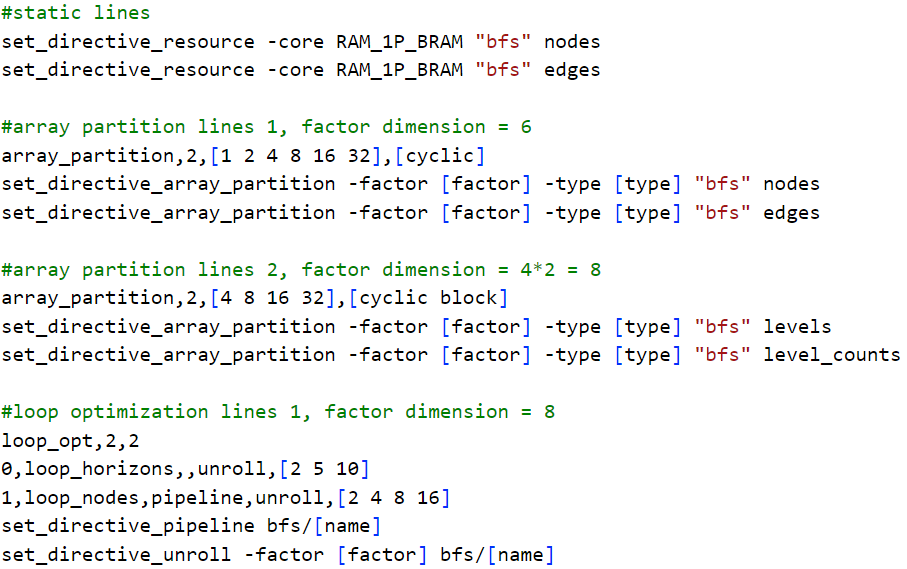}}
\caption{Example template Tcl file to generate the optimization strategy for the application \texttt{bfs} from Machsuite}
\label{fig:example_template}
\end{figure}

An example \textit{template.tcl} is shown in Fig \ref{fig:example_template}. It contains 4 blocks of lines which are classified into three types: static lines, array partition lines and loop optimization lines.

\begin{itemize}[leftmargin=*]
    \item Static lines: The directive lines under static lines are not subject to change, they should be the same and written into every generated Tcl file.
    \item Array partition lines: The first line indicates the sets of parameters applied for HLS. It contains a number denoting the number of directive lines, a list of numbers denoting the factor sets for array partition and a set of types for array partition. The rest of lines are the directive lines with placeholder that should be replaced with the parameters defined by the first line. The placeholders inside the directive lines are replaced with the combination of factor sets and type sets, and every Tcl file will contain one combination of directive parameter. The array partition lines 2 in the Fig \ref{fig:example_template} generates 8 combination directive parameters in this case due to 4 factors and 2 types. Note that factor equalling to 1 means no array partition is applied.
    \item Loop optimization lines: The first line denotes the number of nested loops and the number of directive lines for loop optimization. It is followed by loop optimization parameter lines, each of which indicates the depth of a loop, the name of a loop, whether to apply pipeline to the loop, whether to apply unroll to the loop and unroll factor sets. The rest of the lines are directive lines with placeholders that should be filled with settings from the loop optimization parameter lines. Unroll and pipeline are applied to at most one layer of nested loops, therefore, the number of generated directives is equal to the sum of the number of unroll factors among all the loops and the one without any loop optimization. The loop optimization lines in Fig \ref{fig:example_template} generates 8 combination directives for the loop optimization.
\end{itemize}

The blocks of directive lines are independent of each other, therefore the number of Tcl files is equal to the products of the number of directive parameter combination among all the blocks. In this example template, 384 Tcl files are generated, and different optimization strategies are expected. The method to generate multiple versions of Tcl files is summarized in \textbf{Algorithm 1}, each block of lines will be parsed into an object.

\SetKwComment{Comment}{/* }{ */}
\begin{algorithm}[hbt!]
\DontPrintSemicolon
\small
\caption{Method to generate multiple Tcl files}\label{alg:one}
\SetKwInOut{Input}{Input}
\Input{template.tcl}
\SetKwInOut{Output}{Output}
\Output{$N$ different versions of Tcl files}
$s\_lines, array\_objs, loop\_objs$ from $template.tcl$;\\
Generate $N$ empty Tcl files\;
\Comment{static lines for each Tcl file}
\For{$i \gets 1$ \KwTo $N$}{
    Write $s\_lines$ to Tcl file
}
\Comment{array partition directives}
\ForEach {$o \in array\_objs, f \in o.factors, t \in o.types$}{
    $array\_partition$ with factor $f$ and type $t$\;
    Write $array\_partition$ to Tcl file\;
}
\Comment{loop unroll and pipeline}
\ForEach{$o \in loop\_objs, f \in o.factors$}{
    Get the $loop$ from $loop\_list$ in $o$\;
    Apply $pipeline$ to $loop$ if pipeline applies\;
    Apply $unroll$ to $loop$ with factor $f$ if unroll applies\;
    Write $pipeline$ and $unroll$ to Tcl file\;
}

\label{algo:generation}
\end{algorithm}

\subsection{Data collection}
IR code, IR operator information, FSMD model files from HLS, and resource utilization reports from both HLS and implementation are included in our dataset. In order to get the high-confidence power estimation, we write testbench and run post-implementation functional simulation for vector-based power estimation.

We observe that there is a chance that the HLS tool generates the same design even though different optimization strategies are provided in the Tcl script. This can be caused by aggressive optimization parameters, which are identified as unachievable by the HLS tool. The tool then automatically downgrades the optimization parameters, which can match optimization parameters during another run. Therefore, redundant designs can be generated. We identify redundant designs by checking the hierarchical resource utilization from HLS reports. If two or more designs have exactly the same utilization, only one will be kept in our dataset.

\section{Properties of HLSDataset} \label{Properties}
\begin{figure*}[hbt]
\centering
\includegraphics[width=1\linewidth]{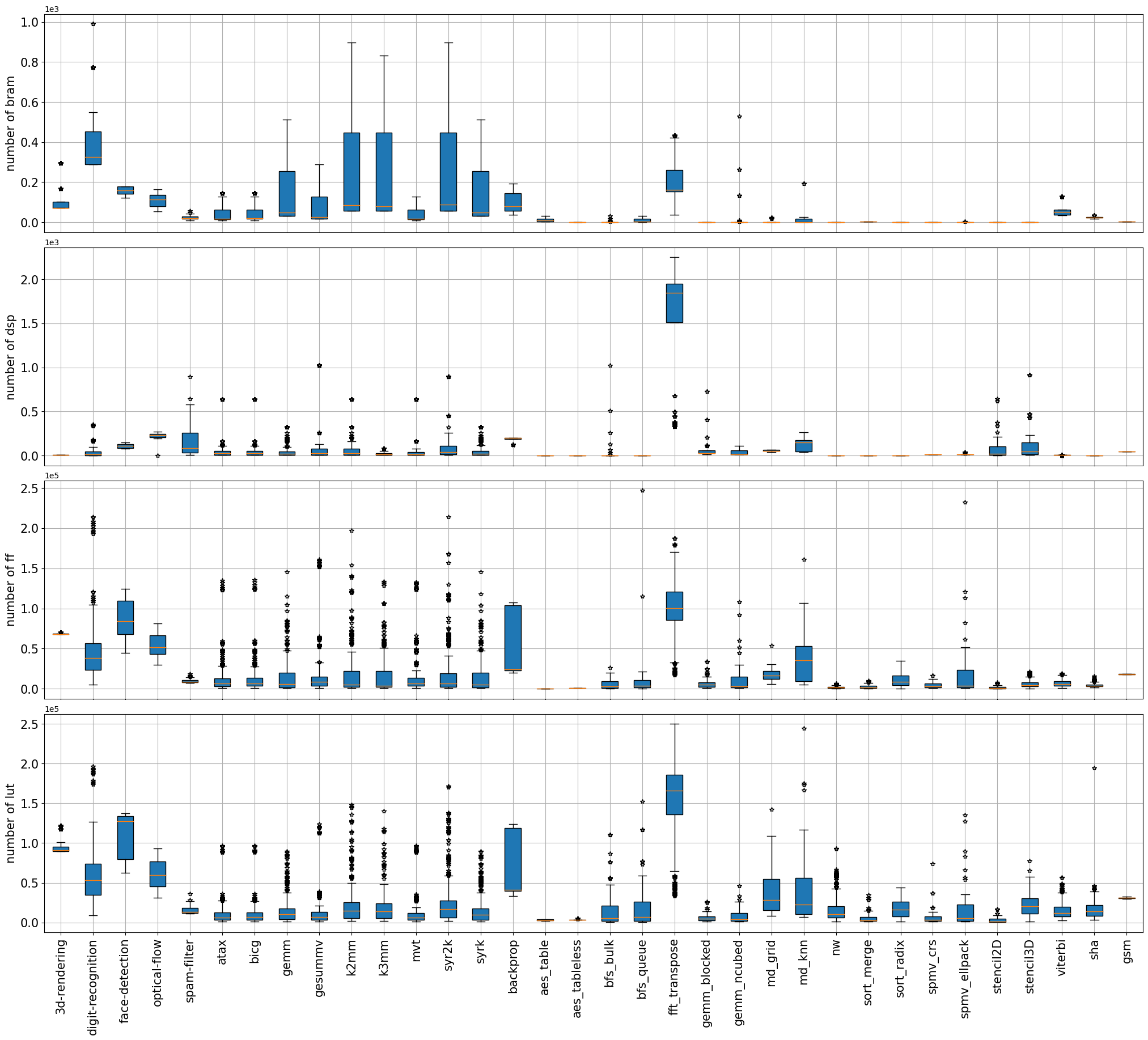}
\caption{Resource utilization of designs generated for ZU9EG, applications are from Rosetta, Polybench, Machsuite and CHStone}
\label{fig:box-plot}
\end{figure*}
\subsection{The contents of HLSDataset}

HLSDataset contains nearly 9,000 hardware design samples for each FPGA type and we consider the features listed below to sufficiently characterize each design sample:
\begin{enumerate}[leftmargin=*]
    \item Resource usage (the number of BRAM, DSP, FF and LUT)
    \item Application domain (e.g., video/graph processing, linear algebra etc)
    \item The number of arithmetic operators (e.g., add, mul), the number of logic operators (e.g., or, shift)
    \item The number/size of primary inputs and outputs
    \item The number of registers, memory and multiplexers
    \item Clock period
\end{enumerate}

Power consumption is also included, since it is crucial when low-power hardware designs are the final target.
We preprocess the raw reports and files from both HLS and implementation phases and generate two CSV files for each benchmark.  Each CSV file contains multiple entries depending on the number of generated hardware designs for the benchmark. The user can directly use the data in the CSV files to train their  ML models, thereby avoiding any effort in changing source code, setting up and running tools, and parsing reports. The detailed contents of the CSV files are listed in Table \ref{table:feature in csv}.

\begin{table}[hbt]
\centering
\captionsetup{font=small}
\renewcommand{\arraystretch}{1.25}
\begin{tabular}{p{1.5cm}  p{6.5cm} } 
\hline
\textbf{Category} &\textbf{\textit{post\_hls\_info.csv} description}\\
\hline
Resource \#     &  Estimated usage and available number of BRAM, DSP, FF and LUT \\
Clock           &  Target, estimation and uncertainty of the clock period \\
Logic ops       &  The number of C and RTL logic operators and associated  resource usage \\
Arith ops       &  The number of C and RTL arithmetic operators and associated resource usage \\
Data ports      &  Width and the number of data input and output ports\\
\hline \hline
\textbf{Category} &\textbf{\textit{post\_implementation\_info.csv} description}\\
\hline
Power           & Simulation-based dynamic power consumption \\
Resource \#     & Actual usage of BRAM, DSP, FF and LUT \\
Clock           & Achieved clock frequency \\
\hline
\end{tabular}
\caption{Descriptions of features included in the CSV files provided with HLSDataset}
\label{table:feature in csv}
\end{table}

It is possible that the features that other researchers are interested in, are not present in the CSV files. Therefore, we also create tar balls containing all the necessary files for feature extraction to do ML training. These files are selected according to how prior works generate their own dataset. Each tar ball contains:
\begin{itemize}[leftmargin=*]
    \item Generated Verilog code (\textit{*.v})
    \item IR code (\textit{*.bc})
    \item IR operator information (\textit{*.adb})
    \item FSMD model (\textit{*.adb.xml})
    \item Resource usage estimation from HLS (\textit{*.verbose.rpt} and \textit{*.verbose.rpt.xml})
    \item Resource utilization reports (\textit{utilization.xls}) and timing reports (\textit{timing.xls}) generated after implementation
\end{itemize}

Considering the reusability and ease-of-use of the dataset, Tcl scripts and source code files are included in the dataset so that researchers can easily extend the dataset with other benchmarks. The detail of how the Tcl script templates can be used is discussed in Section \ref{Construction}. We also include Verilog testbenches so that the generated designs can be easily evaluated with simulation-based power estimation. 

Overall, the contents of HLSDataset are summarized as: 
\begin{enumerate}[leftmargin=*]
    \item The CSV files containing features for each hardware design listed in Table \ref{table:feature in csv}
    \item Tcl templates and actual scripts to generate the dataset
    \item C source code files manipulated with HLS pragmas
    \item Testbenches in Verilog to test generated Verilog designs
    \item Tar balls containing raw files and reports from HLS and reports from implementation stage
\end{enumerate}

Compared to the datasets used by prior works, HLSDataset gives a wider coverage of information for each design, and it gives higher chance for researchers to use or extract useful features directly, meanwhile, no efforts are needed to run the time-consuming HLS and implementation tool flows.




\subsection{Statistical overview of HLSDataset}

Fig \ref{fig:box-plot} provides a view of the diversity of the HLSDataset, through the resource usage metrics of the designs (or samples) contained in the dataset.
We use box and whisker plots to show the distribution of LUTs (Look Up Tables), FFs (Flip Flops), DSPs (Digital Signal Processing Blocks) and BRAMs (Block RAMs) consumed by the designs generated from each application.
As mentioned earlier, we use 4 widely used benchmark sets - Polybench, Machsuite, CHStone and Rosetta - to generate our dataset.
Machsuite and Polybench are mainly composed of short programs and kernels, however, tuning the directives aggressively can still lead to large resource usage on FPGA. Rosetta, on the other hand, is composed of applications from ML and image or video processing domains. Each application of Rosetta contains multiple kernels, and it leads to larger resource usage on FPGA. The secure hash algorithm SHA and linear predictive coding analysis GSM are picked from CHStone due to their representative in the domain. We chose not to include arithmetic operation programs from CHStone due to the limited HLS design space in those applications.

\section{HLSDataset Applications} \label{Applications}
\begin{table*}[hbt]
\centering
\captionsetup{font=small}
\renewcommand{\arraystretch}{1.25}
 \begin{tabular}{|| >{\centering\arraybackslash}m{1cm} >{\centering\arraybackslash}m{1.5cm} >{\centering\arraybackslash}m{2.5cm}  >{\centering\arraybackslash}m{2cm} >{\centering\arraybackslash}m{8.6cm} ||} 
 \hline
 \textbf{Work} &  \textbf{ML model} & \textbf{Task} & \textbf{C source} & \textbf{Feature and source} \\ 
 \hline\hline

\cite{hls-qor} & Lasso, XGB, ANN & Resource usage and timing &  CHStone, Machsuite, S2CBench, Rosetta & Resource usage estimation for logic ops, arithmetic ops, memory and multiplexer; achieved clock period and uncertainty from \textbf{HLS reports}  \\
\hline
\cite{hls-perf} & GNN & Resource usage and timing & CHStone, Machsuite, Polybench & Graph samples based on \textbf{IR code}; operator type, used resource type and timing information from \textbf{HLS reports} \\
\hline
\cite{HL-pow}  &  CNN & Power estimates & Polybench & Resource utilization and clock estimation by \textbf{HLS reports}; signal activities track and IR operator information from \textbf{IR code}; RTL operator information from \textbf{FSMD model} \\
\hline
\cite{PowerGear}  &  GNN & Power estimates & Polybench & Signal activities track and IR operator information from \textbf{IR code}, Graph samples built with \textbf{IR code} and \textbf{FSMD model}, RTL operators information in \textbf{FSMD model} \\
\hline
\cite{dsp-gnn} & GNN & Operation delay & Machsuite & Graph structures, operation type and bitwidth from \textbf{IR code} \\
\hline
\end{tabular}
\caption{Prior ML-based prediction via HLS work, the used ML model, prediction tasks, the used dataset for training and the availability of the dataset.}
\label{table:prior work}
\end{table*}

HLSDataset can be applied to a multitude of prediction applications. 
Table \ref{table:prior work} summarizes the prior works in the area of prediction at the HLS level. The data required for training ML models for each of these prior works is included in HLSDataset. Hence, HLSDataset can be effectively used for these and similar works.

\textbf{Resource utilization estimates:} HLSDataset can be directly used for post-implementation resource utilization estimates. Dai et al. \cite{hls-qor} use Lasso linear model, XGB and artificial neural network (ANN) to improve the quality of HLS-generated resource utilization values with features extracted from HLS reports.
Wu et al. \cite{hls-perf} predict post-implementation resource usage by using the graph structure obtained from the IR codes generated by front-end of HLS tools.
Fast estimation of resource usage find application in design space exploration while generating overlay architectures for FPGAs \cite{overgen}.
The features and feature source used to conduct the studies can be easily found and extracted from HLSDataset.

\textbf{Timing and operation delay prediction: } Wu et al. \cite{hls-perf} demonstrates prediction of post-implementation critical path timing using IR codes and features from HLS reports. D-SAGE~\cite{dsp-gnn} builds graph samples using the IR generated during HLS and use them as input to GNNs to predict operation delay. 
HLSDataset contains the IR code files as well as HLS reports generated by Vivado HLS, and can be used to train such models to predict timing related information. 

\textbf{Power estimates: }HL-Pow \cite{HL-pow} and PowerGear \cite{PowerGear} train ML models to predict power consumption using convolutional neural networks (CNNs) and GNNs respectively. Predicting power consumption needs data such as signal activities and operators obtained from the IR. While those signal activities are not directly included in HLSDataset, testbenches and stimulus are provided so that both RTL-level simulation and C-level simulation can be conducted. Necessary codes to run the simulation: IR codes and RTL designs are included in HLSDataset.

Beyond the above tasks, HLSDataset can be applied to many more potential use cases. While the mentioned works target single-FPGA prediction, HLSDataset includes samples from multiple FPGAs. We believe HLSDataset has the potential to be used for cross-FPGAs metric prediction, although no existing work shows this usecase. 
In addition to prediction of results, HLSDataset can be used to train models to optimize EDA tools and help on faster design space exploration. Moreover, HLSDataset can also be used to evaluate the ML model efficiency in HLS domain with the advancing of ML techniques. 

The features and labels used by each ML models vary widely depending on the task and ML algorithm used, as we can see from table \ref{table:prior work}.
By including information from different levels in the CSV files and TAR balls in HLSDataset, we ensure that all such ML models can be trained.
Researchers can extract information from TAR balls and apply HLSDataset to many other applications.




\section{Case studies} \label{Case studies}
Our dataset covers large amount of features and metrics from post-HLS and post-implementation reports which can be used in machine learning models directly. Therefore, users can simply  extract the necessary information from our dataset to train and test their models. In this section, we perform two case studies by training and testing ML models with HLSDataset to demonstrate the usage of it.

\subsection{Case Study 1: Power Estimation in FPGA HLS via GNNs}

\begin{figure}[hbt]
\centering
\includegraphics[width=1\linewidth]{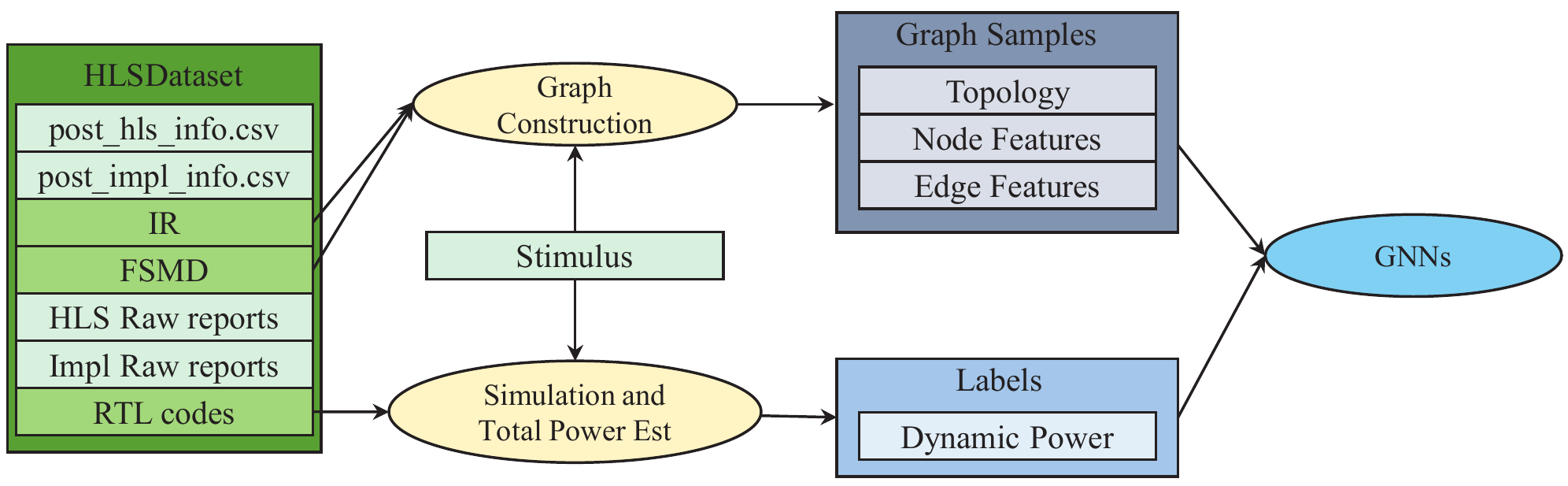}
\caption{Usage of HLSDataset to construct machine learning based power model}
\label{fig:CaseStudy1}
\end{figure}

In our first case study, we replicate the graph neural networks (GNNs) in PowerGear \cite{PowerGear} to predict the post-implementation power using both post-HLS features and signal information extracted from  C-level simulation. We use simulation power as our ground truth power. The GNN models are trained and tested with the subset of HLSDataset on the same FPGA. IR code can be directly used to build graph samples which serve as the inputs to the GNNs. The usage of HLSDataset in this case study is shown in Fig \ref{fig:CaseStudy1}.

The model is trained using the dataset from Polybench. We leave one target application out of the nine applications as the test dataset and use all the rest for training. With the iteration of the approach, we generate one model for every application. We perform 10-fold cross-validation for model generation. All the above steps are repeated for the dataset from the other FPGA. All the training and testing run on Nvidia Ampere A100 GPU. The results for two FPGA devices, ZU9EG and XC7V585T, can be found in Table \ref{table:case study 1}. The test errors for dynamic power range from 3.89\% to 7.93\% on ZU9EG and from 5.25\% to 9.43\% on XC7V585T, and the average errors are 5.08\% and 6.40\% respectively. The results show that HLSDataset can be used to perform ML-based power estimation tasks for FPGA.

\begin{table}[hbt]
\centering
\captionsetup{font=small}
\renewcommand{\arraystretch}{1.25}
\begin{tabular}{|| >{\centering\arraybackslash}p{1.5cm} | >{\centering\arraybackslash}p{2.8cm} | >{\centering\arraybackslash}p{2.8cm} ||}
 \hline
 \multirow{2}{*}{\textbf{Application}} & \multicolumn{2}{c||}{\textbf{Error of Dynamic Power (\%)}} \\
  & \textbf{ZU9EG} & \textbf{XC7V585T} \\
 \hline\hline

atax  &  3.89 &   5.25 \\
\hline
bicg  &  3.90 &   5.60 \\
\hline
gemm &   5.24 &   6.50 \\
\hline
gesummv & 7.93 &  9.43 \\
\hline
k2mm &  4.25 &    6.00 \\
\hline
k3mm  &  4.15 &   6.47 \\
\hline
mvt  &  4.64 &    5.62 \\
\hline
syrk  &  5.31 &   6.22 \\
\hline
syr2k  & 6.41 &   6.46 \\
\hline
\textbf{average} & \textbf{5.08} & \textbf{6.40} \\
\hline
\end{tabular}
\caption{\textbf{Dynamic power estimation errors} - Training dataset and testing dataset are from Polybench subset of HLSDataset. Results for ZU9EG and XC7V585T.}
\label{table:case study 1}
\end{table}

\subsection{Case Study 2: Estimation of Quality of Results in HLS with ML}

\begin{figure}[hbt]
\centering
\includegraphics[width=1\linewidth]{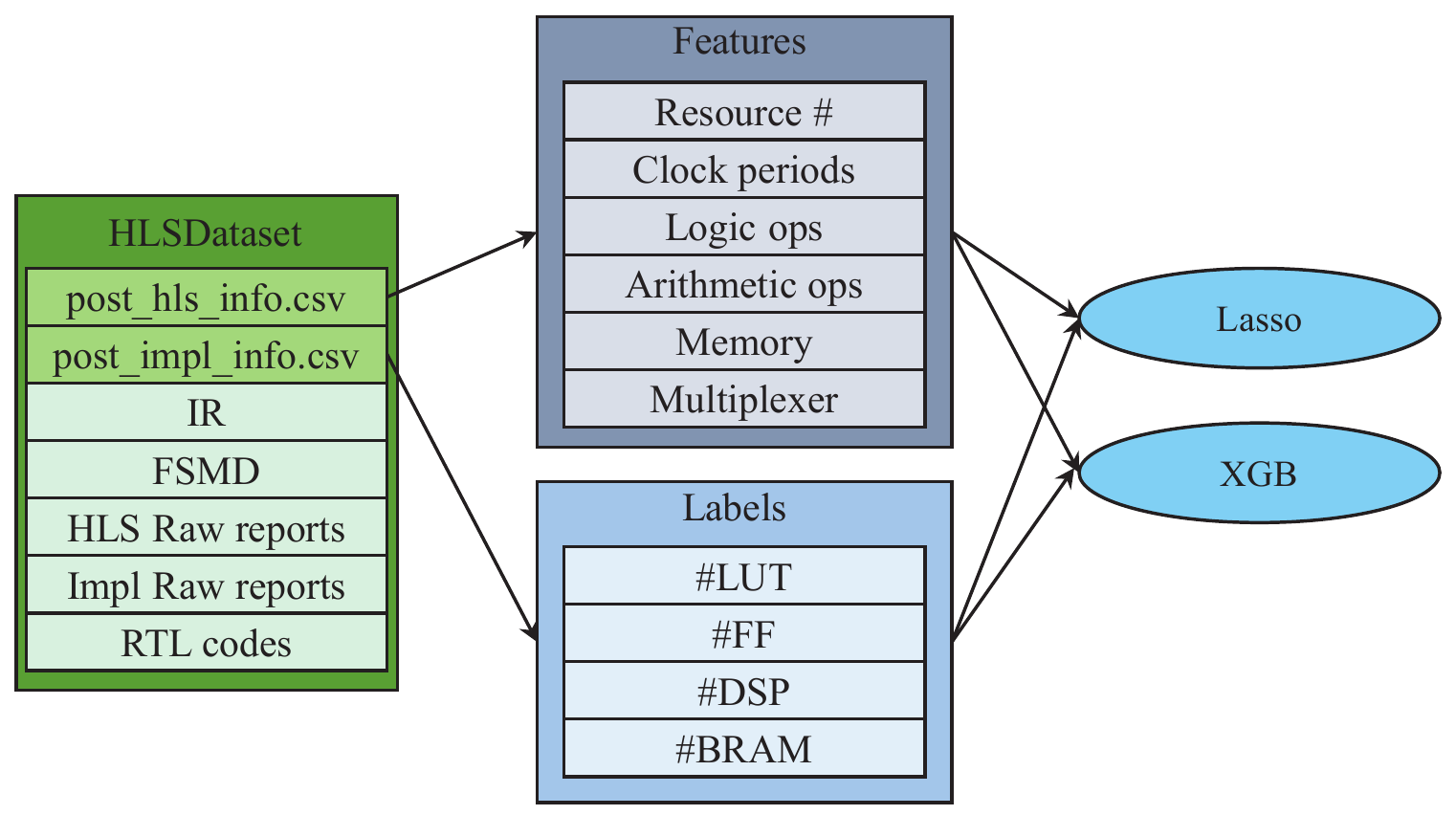}
\caption{Usage of HLSDataset to construct machine learning model for estimation of resource utilization}
\label{fig:CaseStudy2}
\end{figure}

The resource usage estimation (LUTs, FFs, DSPs, BRAMs) generated by HLS tools are fast but inaccurate compared to the post-implementation reports because HLS tools simply sum up the contributions of instantiated functional units during the synthesis. This approach fails to capture the optimization effects and limitations imposed by resources on-chip. However, as S. Dai et al. \cite{hls-qor} indicates, ML can help to predict more-accurate resource usage from estimates in the HLS reports. 

We replicate the ML model but use our HLSDataset as training and test set to evaluate the ML model on estimation of post-implementation resource usage. The way to use HLSDataset is illustrated in Fig \ref{fig:CaseStudy2}.
Machsuite, Polybench subsets from HLSDataset are used to train the XGB and Lasso linear model. The features are extracted from FSMD file (.adb.xml) and resource estimates reports (.verbose.rpt.xml). The ground-truth resource utilization is extracted from post-implementation reports. All the files and reports are included in our dataset, only a parser is needed to extract necessary data to be used in the ML model. Single-task XGB and Lasso model are used in our case. We randomly select 20\% of 8735 samples from the subsets as the testing set and the rest as the training/validation test set. 10-fold cross-validation is performed during training, and 75\% of the training/validation set is selected for training and 25\% for validation. The results are shown in Table \ref{table:case study 2}. The HLS tool fails to provide good estimates for LUT and FF usage, while DSP and BRAM estimates are accurate. XGB and Lasso demonstrate a significant accuracy improvement in the estimation of LUT and FF usage. The results shown in this table differ from those in the original paper because there are differences in target FPGA, the dataset, features used to train the model and the version of HLS tools used for the dataset generation. Therefore, we do not show a comparison with the original work here.

\begin{table}[hbt]
\centering
\captionsetup{font=small}
\renewcommand{\arraystretch}{1.25}
\begin{tabular}{|| >{\centering\arraybackslash}p{1.8cm} | >{\centering\arraybackslash}p{1.1cm} | >{\centering\arraybackslash}p{1.1cm} 
|>{\centering\arraybackslash}p{1.1cm} | >{\centering\arraybackslash}p{1.2cm} ||}
\hline
\textbf{Resource} & \textbf{LUT} & \textbf{FF} & \textbf{DSP} & \textbf{BRAM} \\\hline
\textbf{HLS Estimate} & 63.2\% & 34.1\% & 0.0\% & 1.8\% \\\hline
\textbf{XGB} & 3.2\% & 2.3\% & NA & 0.1\% \\\hline
\textbf{Lasso} & 13.2\% & 15.4\% & NA & NA \\\hline
\end{tabular}
\caption{\textbf{Resource estimation errors} - Training dataset and testing dataset are from Machsuite and Polybench subsets of HLSDataset. Results for ZU9EG.}
\label{table:case study 2}
\end{table}
\section{Conclusion} \label{Conclusion}
This work presents HLSDataset, a dataset for ML-assisted FPGA design using HLS. HLSDataset covers a wider range of data than other datasets in this domain, and is the first open-source dataset of its kind that can be used for multiple studies.
We demonstrate that HLSDataset can be used in training ML models targeting different applications such as resource usage prediction, power prediction, etc. We also present the methodology to generate the dataset so that HLSDataset can be futher extended. 

We are currently expanding HLSDataset by including data for more target frequencies (e.g. clock period = 5ns, 2.5ns, etc.).
For future work, we plan to extend HLSDataset to include more benchmarks (e.g., S2CBench) 
and more FPGAs (including Intel FPGAs).
While the design samples in HLSDataset are generated from C benchmark so that ML-assisted HLS based studies can be conducted, we plan to extend the dataset to include data from native Verilog benchmarks so that ML-assisted Verilog based studies are possible with our dataset.
\section{Acknowledgement}
We thank all the anonymous reviewers for the detailed comments on the paper. This work was supported in part by the AI4AI award from Meta and the National Science Foundation grant 1725743. Any opinions, findings, conclusions, or recommendations are those of the authors and do not necessarily reflect the views of these funding agencies. 

\bibliographystyle{IEEEtran}
\bibliography{9_reference}




%

\end{document}